\documentclass{article}

\usepackage{arxiv}

\usepackage[utf8]{inputenc} 
\usepackage[T1]{fontenc}    
\usepackage{hyperref}       
\usepackage{url}            
\usepackage{booktabs}       
\usepackage{amsfonts}       
\usepackage{nicefrac}       
\usepackage{microtype}      
\usepackage{lipsum}
\usepackage{graphicx}
\graphicspath{ {./images/} }

\title{The role of extreme geomagnetic storms in the Forbush decrease profile }

\author{Anil Raghav$^{1}$, Prathmesh Tari$^{1}$, Kalpesh Ghag$^{1}$,
Zubair Shaikh$^{2}$,Omkar Dhamane$^{1}$,Utsav Panchal$^{1}$,\\
\textbf{Mayuri Katvankar$^{1}$, Komal Choraghe$^{1}$, Digvijay Mishra$^{1}$, and Kishor Kumbhar$^{1}$}\\ \\
$^{1}$Department of Physics, University of Mumbai, Mumbai, India\\
$^{2}$Indian Institute of Geomagnetism (IIG), New Panvel, Navi Mumbai-410218, India\\
}

\begin{document}
\maketitle
\begin{abstract}
The Forbush decrease (FD) and Geomagnetic storm (GS) are the two distinct space weather events having common causing agents like interplanetary coronal mass ejection (ICME) or corotating interacting region (CIR). Generally, an ICME causes high amplitude FDs and extreme GSs.  However, the interlinks between extreme GS and strong FDs are poorly studied. Here, we demonstrate five ICME induced extreme storms and their effects on respective FD profiles. We observed the sudden storm commencement of the GS coincides with the FD onset. Interestingly, we also noted a gradual increase in neutron counts during the main and recovery phases of GS.  The maximum enhancement in neutron counts coincides with the minimum value of the Sym-H index. The enhancement is visible primarily in all the neutron monitors but significantly pronounced in high-energy neutrons compared to low-energy neutrons. The weakening of Earth's magnetic shield due to  ICME-Magnetosphere interaction allows more cosmic rays to reach the ground.  Thus, we conclude that the geomagnetic storm conditions highly influence the FD profile along with the external causing agent. Therefore, it is essential to include the effect of geomagnetic field variation in the models that are used to reproduce the FD profile.  
\end{abstract}


\section{Introduction}

The sudden short-time decrease in cosmic-ray flux followed by recovery is known as the Forbush decrease (FD) phenomena \cite{forbush1937effects,hess1937world}. The decrease occurs for a few hours, whereas recovery lasts from several hours to days and sometimes even weeks. FDs plays an important role in different atmospheric process such as; (i) cloud and aerosol physics \cite{svensmark2016response}, e.g.,  high amplitude FDs ($>7\%$) significantly affects the diurnal temperature range  \cite{dragic2011forbush}; (ii) statistical study suggest that FDs main phase causes 17\% decrease in rainfall over the USSR \cite{stozhkov1995rainfalls,stozhkov1996effect,stozhkov2003role}; (iii) FDs causes 18\% to 12\% reduction of cloud cover \cite{todd2001changes,todd2004short}, Etc. Moreover, FDs are also observed on Mars surface as well as on the Martian atmosphere \cite{guo2018measurements,witasse2017interplanetary,papaioannou2019catalogue,freiherr2019tracking}, and lunar space environment \cite{sohn2019forbush}.   Furthermore, FD measurement at  Mercury, Earth, and Mars suggested that FD size decreases exponentially with increasing heliocentric distance \cite{winslow2018opening}. Thus, FDs are global phenomena and play a crucial role in the planetary atmosphere.


The FDs are generally induced either by interplanetary coronal mass ejections (ICMEs) or by corotating interaction regions (CIRs). Thus, Forbush decrease events are divided into two basic categories; (a) non-recurrent FDs caused by ICMEs and (b) recurrent FDs caused by CIRs \cite{belov2014coronal}. The first type shows a sudden onset, attain maximum decrease within about a day, and have a gradual or fast recovery. On the other hand, the second type has a more gradual onset and decrease with almost similar recovery duration that leads to symmetric FD profile. Moreover, ICMEs have two sub-structures named as a sheath (along with shock) and magnetic cloud (MC) \cite{zurbuchen2006situ}. Therefore, ICMEs leads to one or two-step FD profile, depending on the transit of the sub-structures to the observer. If an only sheath or MC crosses to the observer, we observe one step FD profile \cite{richardson2011galactic}. Whereas, if both sheath and MC pass the observer, then we observe a two-step FD profile; the first step is induced by the shock (including sheath), and the later step is caused by MC  \cite{raghav2014quantitative,cane2000coronal,arunbabu2013high}. The reported correlation studies \cite{yadav1986influence,belov1997forbush,belov2001determines,belov2008forbush,belov2014coronal,dumbovic2011cosmic,dumbovic2012cosmic,richardson2011galactic,richardson2010near} have noted a significant correlation between the FD amplitude and associated magnitude and duration of the interplanetary magnetic field and solar wind speed. The relative comparisons with white-light coronagraphic observations have suggested the FD magnitude to be larger for (i) faster CMEs \cite{belov2008forbush,blanco2013observable,belov2014coronal}, (ii) CMEs with larger apparent width \cite{kumar2014cosmic,belov2014coronal}, and (iii) CMEs with greater mass \cite{belov2014coronal}. \cite{raghav2017forbush} studied 16 ICME induced Forbush decrease events of large magnitude and emphasizes the complexity observed in FD profiles during shock-sheath and MC transit, e.g., one step or multi-step decrease, simultaneous/non-simultaneous decrease with respect to the shock-front/MC arrival at the Earth, a gradual decrease or short-duration sharp decrease, Etc. In fact, recent studies suggested the influence of small-scale magnetic structure associated within the ICME  sheath on cosmic-ray variations \cite{jordan2011revisiting, shaikh2017presence,shaikh2018identification}. They suggested that small-scale flux-rope and planar magnetic structure-like structure can evolve within an ICME sheath, which causes local recovery or local decrease in FD profile.

Furthermore,  cosmic ray modulation was investigated using  a transport equation that includes the diffusion, gradient, and curvature drifts arising because of the magnetic field, convection-diffusion of the solar wind, and energy change due to compression or expansion of the fluid \cite{parker1965passage,potgieter2013solar}. Various theoretical models were proposed to explain the features of the FD profile such as diffusion \cite{kota1990diffusion,kota1991role}, diffusion-convection \cite{gleeson1968solar,richardson1996relationship,kota2013theory,bhaskar2016relative,raghav2020exploring}, guiding center drift \cite{jokipii1981effects,kota1983effects,krittinatham2009drift} and energy change \cite{potgieter1987radial,jokipii1989physics} form the basis for the suggested models. In addition, models based on (i) ordered and/or turbulent magnetic fields in the interplanetary medium, (ii) convection and adiabatic energy loss by a fast stream, (iii) enhanced drift as well as scattering properties of a strong and fluctuating magnetic field, (iv) effect of finite Larmor radius, etc. \cite{lockwood1986characteristic,yadav1986influence,zhang1988magnetic,badruddin2002transient,morrison1956solar,wawrzynczak2010modeling,candia2004diffusion,alania2008forbush,subramanian2009forbush,wibberenz1997two,wibberenz1998transient,arunbabu2013high,raghav2014quantitative,le1991simulation,wibberenz1997two,wibberenz1998transient,krittinatham2009drift,kubo2010effect} are also deliberated in literature to understand the FD. However, non of the models were able to reproduce the complete profile of different FDs.

Similar to FDs, the Geomagnetic storms (GSs), which are a temporary disruption in the Earth's magnetosphere, have a common solar and interplanetary origin in the form of ICMEs or CIRs \cite{kamide1998current,richardson2012solar,akasofu2018review,kumar2015study}. ICME induced GS also has different profiles such as; single step, two-step, fast recovery, slow recovery, fast recovery followed by a slow recovery, etc \cite{raghav2019cause,shaikh2019concurrent,raghav2019cause}. The decreasing phase is associated with the ring current intensification, whereas the recovery phase is due to the decay of ring current \cite{akasofu2018review,jordanova2020ring}. They proposed that ring current can decay due to charge exchange mechanism, wave-particle interactions, increase in recombination rate, Etc. The storm's strength is denoted with the index $Dst$ or $Sym-H$. The most important parameter for the GS to occur is the southward interplanetary magnetic field or convective electric field.  Since the causing agent of FDs and GSs are the same, therefore, the association of FDs with GSs and solar wind disturbances is a subject of continuous interest \cite{yoshida1966development,yoshida1968ring,baisultanova1995magnetospheric,belov2005magnetospheric,belov2008forbush,kane2010severe,kumar2015study,lingri2016solar,okike2021testing}.


\section{Data and Method}


This study uses the neutron flux data measured by various Neutron Monitor (NM) observatories as a proxy of cosmic ray flux. The NM data is available at the worldwide neutron monitor database \footnote{\url{https://www.nmdb.eu/nest/}}. The ground-based measurement of cosmic ray flux depends on the local rigidity, altitude, latitude, and longitude. Therefore, it is essential to include baseline correction for comparative study. It is essential to normalize the measured neutron flux data with quiet condition data considering variations  in each neutron monitor observatory. The normalized percentage variation ($\%$) for each neutron monitor observatory is defined as;

\begin{equation}
	N_{norm}(t)= \frac{N(t)-N_{mean}}{N_{mean}} \times 100
\end{equation}
where $N_{mean}$ is averages of quiet day/days, neutron flux of a specific observatory, and N(t) is neutron flux at time t of the same specific observatory. We have used $5~min$ temporal resolution of the neutron flux data for our analysis. It is essential to understand the cosmic ray modulation based on the energy of cosmic ray particles. Therefore, we classify the neutron monitor data into three broad energy bands, (i) low rigidity (0–2 GV), (ii) medium rigidity (2–5 GV), and (iii) high rigidity ($\ge$ 5 GV).

Furthermore, to study the magnetospheric disturbed conditions, we have utilized geomagnetic storm index i.e., Sym-H index. For the analysis of interplanetary structures that cause the disturbance in magnetosphere, we have used data from OMNI database \footnote{Available at \url{https://cdaweb.gsfc.nasa.gov/}}. It provides interplanetary plasma and magnetic field data which is time shifted data at Earth's bow-shock nose. The interplanetary magnetic field (IMF) ($B_T (nT)$), and their components $B_x$, $B_y$, $B_z$ (nT) in GSM coordinate system. The plasma parameters includes, solar wind speed ($V_p (kms^{-1})$), and plasma temperature ($T (K)$), proton density ($N_p (cm^{-3})$), flow Pressure($P (nPa)$), Electric Field($E (mV/m)$), and plasma beta ($\beta$). Both, the Sym-H index, plasma parameters, and magnetic field data has $1~min$ temporal resolution.

\section{Observations}
We have analysed 31 extreme storm events ($Sym-H \leq 200 ~nT$) listed in \cite{choraghe2021properties}, to study the role of disturbed geomagnetic conditions in the respective FDs. In most extreme GS events, we observed recovery or spike in the neutron flux data associated with the minimum Sym-H value. It is important to note that the GS and FD exhibit complex profiles. Therefore, it isn't easy to establish a strong link between Sym-H and FD index. Thus, in many events, the signature is not explicitly distinguishable. Here, we are demonstrating five extreme geomagnetic storm events to conclude the interlinks between GS and FD. The selected events are (i) 26 August, 2018; (ii) 20 November, 2003; (iii) 24 November, 2001 (iv) 10 May, 1992 and (v) 13 July, 1991. The event-wise description is provided as below;

\subsection{26 August 2018}
\begin{figure}
	\includegraphics[width=\columnwidth]{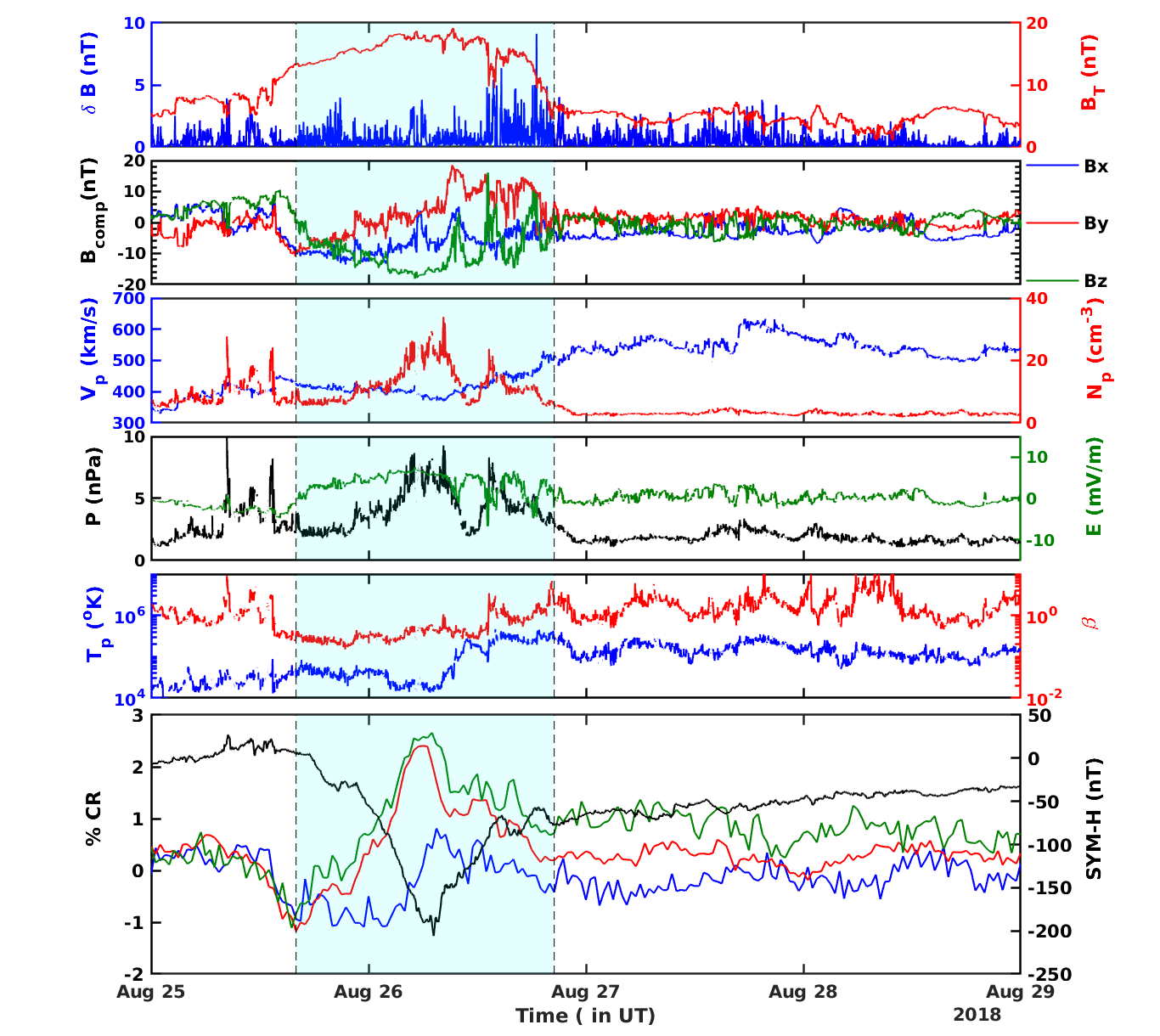}
	\caption{Top five panels show OMNI database of interplanetary parameters during ICME passage from 25-30 August 2018. The topmost panel display total interplanetary field strength IMF ($B_T$) and variation in IMF i.e. $\delta B$. The 2nd panel from top show temporal variations in IMF components ($B_x$, $B_y$, $B_z$,). The third panel demonstrate speed of solar wind ($V_p$) and proton density ($N_p$). The Fourth panel indicates plasma flow Pressure ($P$) and Electric Field ($E$). The fifth panel shows plasma temperature ($T_p$) and plasma beta($\beta$). The bottom panel indicated the variations in normalized cosmic ray  (CR) flux and Sym-H index. In bottom panel, (i) low rigidity (0–2 GV) CR flux indicated by blue line, (ii) medium rigidity (2–5 GV) CR flux indicated by red line, and (iii) high rigidity ($\ge$ 5 GV) CR flux indicated by green line. The cyan shaded region indicates the enhancement in the cosmic ray flux and its association with Sym-H index.}
	\label{fig:2018}
\end{figure}

The extreme storm with SYM-H $\sim -200 (nT)$ occurred on 26 August 2018. The interplanetary parameters,  geomagnetic index, and neutron flux variations are demonstrated in Figure \ref{fig:2018}. The cyan shed represent the transit of the ICME MC during 25-26 August, which shows enhancement in a magnetic field ($B_T$) along with smooth rotation in its components ($B_x$, $B_y$, $B_z$), lower plasma beta, lower temperature, etc. The details of the MC characteristics can be found in \cite{piersanti2020sun,kihara2021peculiar}. During the transit of MC, we observe that it contributes to the main phase and fast recovery phase of the storm.  The cosmic ray flux decrease started in the initial phase of the storm and simultaneously with the onset of the magnetic cloud. The FD shows a decrease of $\sim 1.4 \%$. Moreover, we observed recovery or enhancement of cosmic ray flux during the main phase of GS. We observed approximately 3.7 \% of enhancement in cosmic ray flux for intermediate and high energy neutron monitors. It is visible that recovery of cosmic rays coincide with the decreasing phase of the storm. Interestingly, maximum cosmic ray flux goes along with the minimum of the Sym-H index value.  Furthermore, the cosmic ray flux decreases as a Sym-H index value recovers. Thus, the enhancement in the cosmic ray flux directly relates to the decreasing phase of the geomagnetic storm.

\subsection{20 November 2003} 
\begin{figure}
	\includegraphics[width=\columnwidth]{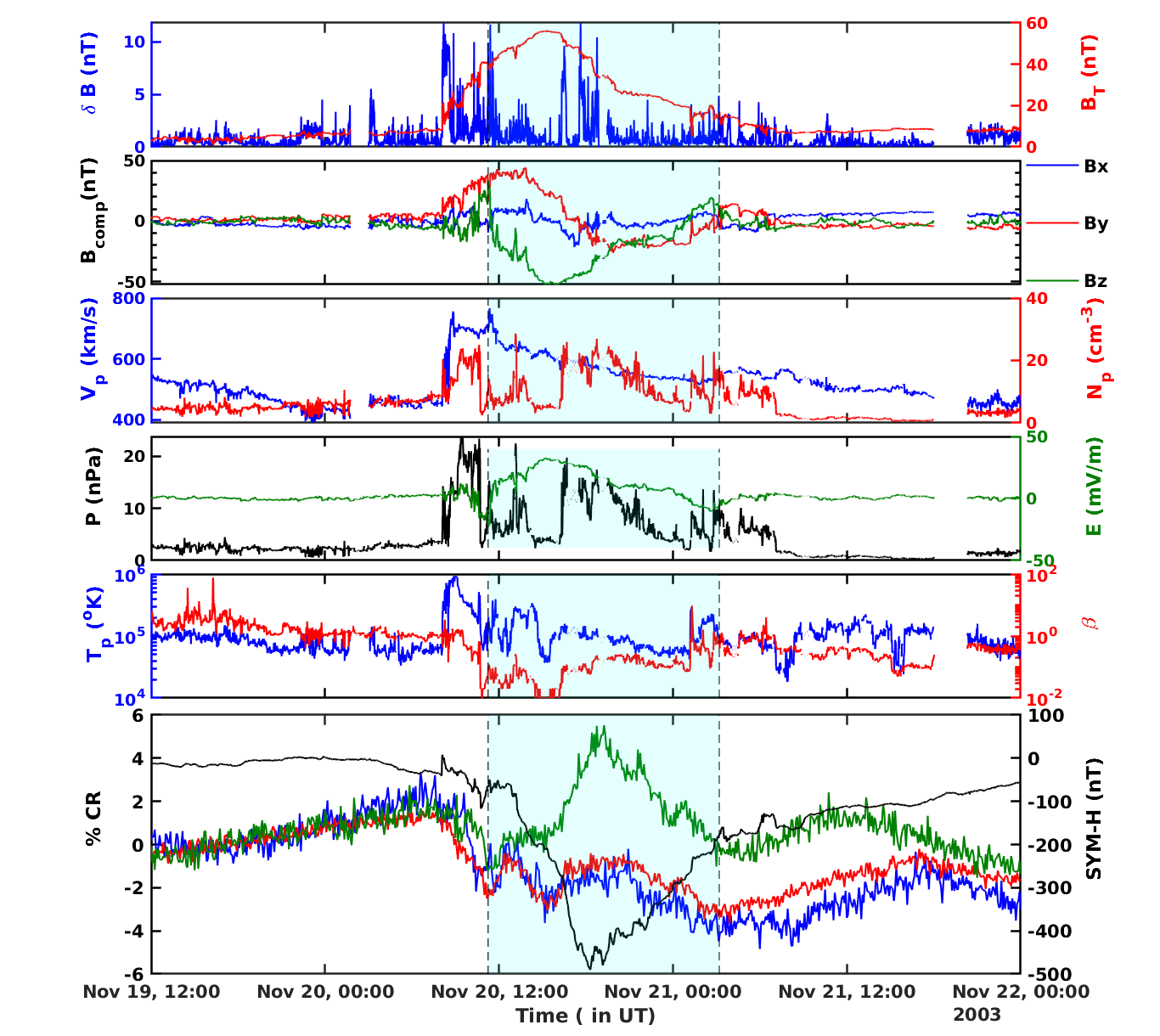}
	\caption{Top five panels show OMNI database of interplanetary parameters during ICME passage from 20-23 November 2003. The topmost panel display total interplanetary field strength IMF ($B_T$) and variation in IMF i.e. $\delta B$. The 2nd panel from top show temporal variations in IMF components ($B_x$, $B_y$, $B_z$,). The third panel demonstrate speed of solar wind ($V_p$) and proton density ($N_p$). The Fourth panel indicates plasma flow Pressure ($P$) and Electric Field ($E$). The fifth panel shows plasma temperature ($T_p$) and plasma beta($\beta$). The bottom panel indicated the variations in normalized cosmic ray  (CR) flux and Sym-H index. In bottom panel, (i) low rigidity (0–2 GV) CR flux indicated by blue line, (ii) medium rigidity (2–5 GV) CR flux indicated by red line, and (iii) high rigidity ($\ge$ 5 GV) CR flux indicated by green line. The cyan shaded region indicates the enhancement in the cosmic ray flux and its association with Sym-H index.}
	\label{fig:2003}
\end{figure}

On $20^{th}$ November 2003, the strongest storm of the current century was observed with Sym-H $\sim -475 (nT)$. Figure \ref{fig:2003} display interplanetary parameters, disturbed geomagnetic conditions, and neutron flux variations for this event. The sudden enhancement in IMF, solar wind speed, plasma density, flow pressure, beta value indicate the onset of the ICME shock front. 
The smooth rotation of magnetic field components and low plasma beta value suggested the signature of magnetic cloud passage during 20-21 November \cite{raghav2020pancaking}. The decrease in cosmic ray flux during the initial phase of a storm is observed with the ICME onset. The FD shows a decrease of $\sim 3 \%$.
Furthermore, a recovery or enhancement of cosmic ray flux is observed during the main phase of GS. It indicates approximately $\sim 7 \%$ of enhancement in cosmic ray flux for high energy neutron monitors. The enhancement is minor for intermediate and low energy cosmic ray flux. The gradual increase in high-energy cosmic ray flux coincides with the storm's main phase. The maximum cosmic ray flux is accompanied by the minimum of the Sym-H index value. Subsequently, the cosmic ray flux decreases as the Sym-H index value recovers. Thus, observation suggests that the high-energy cosmic ray flux enhancement is directly associated with the GS's maximum amplitude.

\subsection{24th November 2001}
\begin{figure}
	\includegraphics[width=\columnwidth]{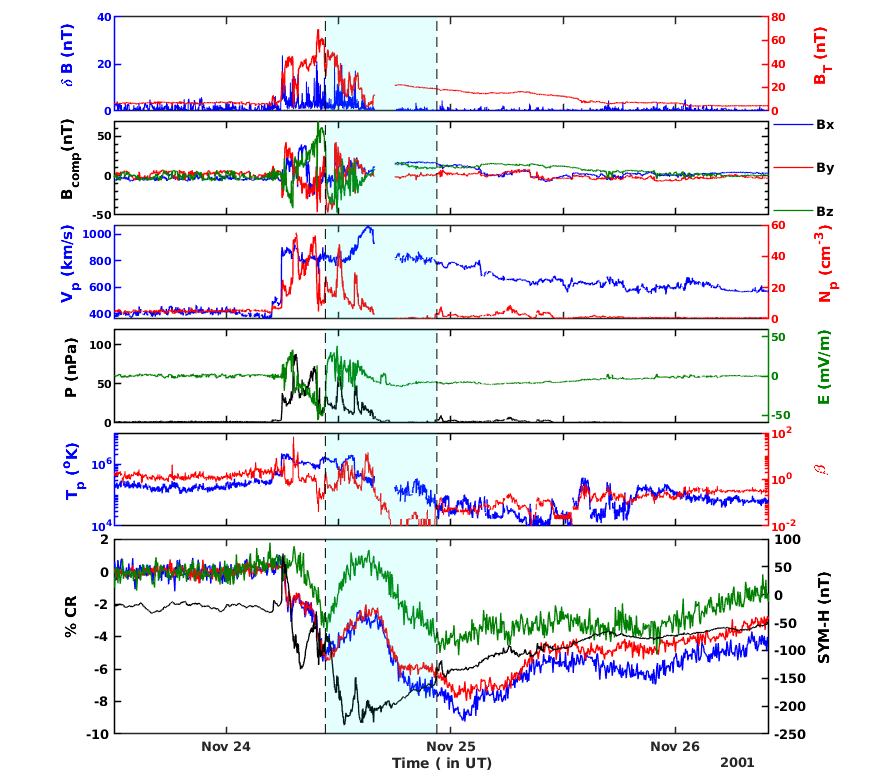}
	\caption{Top five panels show OMNI database of interplanetary parameters during ICME passage from 24-2930 November 2001. The topmost panel display total interplanetary field strength IMF ($B_T$) and variation in IMF i.e. $\delta B$. The 2nd panel from top show temporal variations in IMF components ($B_x$, $B_y$, $B_z$,). The third panel demonstrate speed of solar wind ($V_p$) and proton density ($N_p$). The Fourth panel indicates plasma flow Pressure ($P$) and Electric Field ($E$). The fifth panel shows plasma temperature ($T_p$) and plasma beta($\beta$). The bottom panel indicated the variations in normalized cosmic ray  (CR) flux and Sym-H index. In bottom panel, (i) low rigidity (0–2 GV) CR flux indicated by blue line, (ii) medium rigidity (2–5 GV) CR flux indicated by red line, and (iii) high rigidity ($\ge$ 5 GV) CR flux indicated by green line. The cyan shaded region indicates the enhancement in the cosmic ray flux and its association with Sym-H index.}
	\label{fig:2001}
\end{figure}

We observe another extreme storm event on $24^{th}$ November 2001 with Sym-H $\sim -230 (nT)$. The interplanetary parameters, disturbed geomagnetic conditions, and neutron flux variations for this event are demonstrated in Figure \ref{fig:2001}. The visible signs of ICME passage are observed during 24-26 November. \cite{falkenberg2011multipoint, raghav2017forbush} 
The sudden enhancement in IMF total, solar wind speed, plasma density, flow pressure, temperature, and beta value indicate the onset of shock-front. It is followed highly turbulent sheath region and magnetic cloud region. The decrease in cosmic ray flux and the onset of magnetic cloud coincides in the initial storm phase. The first step of FD shows a decrease of $\sim 5 \%$. We observed recovery or enhancement of cosmic ray flux during the main phase of GS. It indicates approximately $\sim 5 \%$ of enhancement in cosmic ray flux for high energy neutron monitors. The enhancement in intermediate and low energy fluxes are approximately $\sim 3 \%$. Similar to a previous couple of events, we observed a gradual increase in cosmic ray flux associated with the gradual decrease of Sym-H index during the main phase.  The maximum cosmic ray flux coincides with the minimum of the Sym-H index value. Thus, it seems that the enhancement in the cosmic ray flux directly relates to the geomagnetic storm peak.

\subsection{13 July 1992 and 10 May 1991}
\begin{figure}
	\includegraphics[width=\columnwidth]{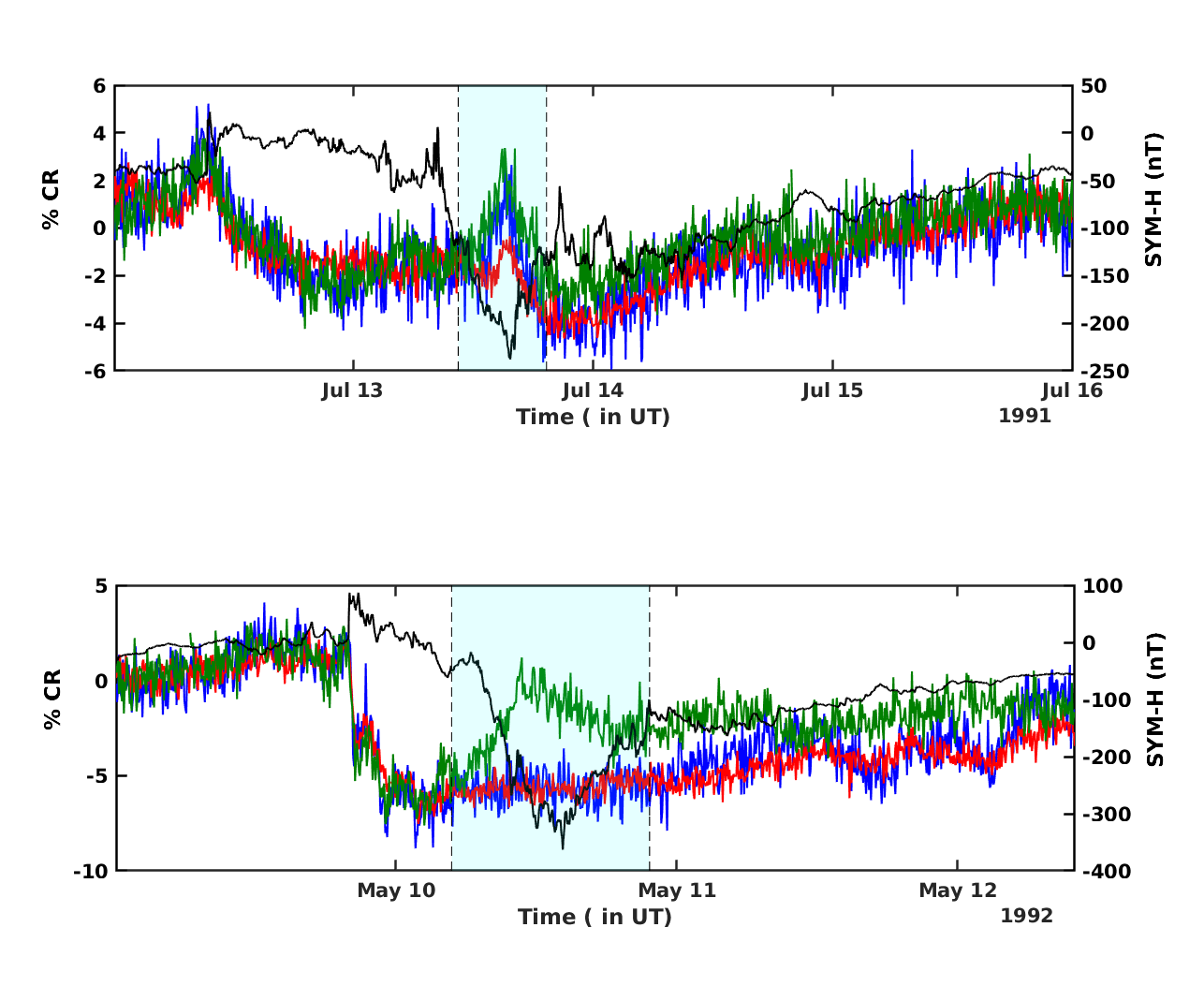}
	\caption{The top and bottom panel show the CR flux and Sym-H variation during 10-13 May 1992 and 13-16 July 1991 events respectively. In both panels, (i) low rigidity (0–2 GV) CR flux indicated by blue line, (ii) medium rigidity (2–5 GV) CR flux indicated by red line, and (iii) high rigidity ($\ge$ 5 GV) CR flux indicated by green line. The cyan shaded region indicates the enhancement in the cosmic ray flux and its association with Sym-H index. }
	\label{fig:1991-92}
\end{figure}

We have also analyzed the extreme storm events that occurred on $10^{th}$ May 1992 and $13^{th}$ July 1991 \cite{balogh1993evolution}. The interplanetary observations of these events are not available. For both events, disturbed geomagnetic conditions and neutron flux variations are demonstrated in Figure \ref{fig:1991-92}. On 10 May 1992, the neutron monitor flux decreased $\sim 7 \%$ whereas for 13 July 1992 event; we observed the FD of $\sim 3 \%$. For both events, we observed recovery or enhancement of cosmic ray flux during the main phase of GS. It indicates approximately $\sim 8 \%$ of enhancement in cosmic ray flux for high and intermediate energy neutron monitors on 10 May 1992. Whereas approximately $\sim 6 \%$ of enhancement in cosmic ray flux for high, intermediate, and low energy neutron monitors on 13 July 1992. We observed a gradual increase in cosmic ray flux and a decrease in Sym-H index for both events.  The maximum cosmic ray flux goes along with the minimum of the Sym-H index value.

From above 5 events we observed that the enhancement in cosmic ray flux during Forbush decrease is contributed by magnetospheric disturbance of geomagnetic storm.

\section{Discussion and Conclusion}        

 The FD profile investigations have been started in the decade of 1940s. The several decades of the study suggest the FD phenomenon is highly complicated due to dynamic space weather conditions in interplanetary space. The reported studies suggested that the variations in cosmic ray fluxes are associated with the solar wind speed, turbulence, and magnetic field intensity. 
 The FD profiles show huge event-to-event variations. Thus, the observations will infer which processes may be more important than others. Recently, \cite{raghav2020exploring} provided a unified approach considering on solar wind speed-based diffusion-convection model to explain the FD originating from ICME and CIR. The turbulence and magnetic field diffusion models also pointed out that the two steps are associated with the ICME sheath and magnetic cloud \cite{arunbabu2013high,raghav2014quantitative,bhaskar2016relative}. The present study aims to identify the role of extreme storm conditions in the evolution of FD profiles.

For the first time,  \cite{yoshida1966development} studied GS and FD phenomena together. They suggested that GS main phase and FD are not related to each other. They also proposed that the main phase of the FD is caused by the turbulent region behind the barrier. \cite{yoshida1968ring} noted an increase of cosmic-ray flux superposed on the FD profile, which could be associated to the intensity of the ring current generated field. They also suggested that the increase is slight or absent for weak GS irrespective of the FD magnitude. Moreover, \cite{baisultanova1995magnetospheric} studied cosmic ray variations associated with changes of cut-off rigidity during the large GS. Still, no relations were found to claim inter-linkages between FD and GS profiles.  Further \cite{kane2010severe} investigated six largest GS and six largest FD and suggesting that large Dst avoids large FDs (and vice-versa). They also pointed that maximum negative Dst and maximum FD amplitude did not occur at the same hour. \cite{badruddin2019forbush} performed cross-correlation analysis between cosmic ray intensity variation with that of the geomagnetic index and suggested a delay of the storm by 3-4 hours. This indicates that FD can provide storm information by one to a few hours scale in advance. \cite{adhikari2021spectral} also noted a strong correlation delay response between FD and storm. They further suggested that the storm index should be use to predict FD. Thus interlinks between FD phenomena and geomagnetic storms are debatable till date. Therefore, it is essential to examine the role of global magnetospheric disturbance  into the FD profile.

This study demonstrates five typical extreme GS events and associated FD profiles. The observations of IMF measurement and interplanetary plasma parameters suggest that three events are associated with ICMEs. Unfortunately, the data is unavailable for the remaining two events, but reported literature indicated that ICMEs generally trigger the extreme GS. So, we conclude that the demonstrated typical events are ICME induced events. The observation shows that the percentage decrease in cosmic ray intensity is initiated with the onset of ICME. The first step of FD is visible in the initial phase of the GS. Subsequently, a gradual increase in cosmic ray intensity is seen in all the events. The first two typical events on 26 August 2018 and 20 November 2003 show significant enhancement above the quiet cosmic ray intensity.
In contrast, remaining events display recovery up to background cosmic ray intensity. Interestingly, the maximum enhancement of cosmic ray intensity is nearly matched with the minimum of Sym-H index value. That implies that the highest comic ray recovery is associated with the highest GS amplitude. Furthermore, a gradual decrease is observed during the recovery of the Sym-H index.

\begin{figure}
	\includegraphics[width=\columnwidth]{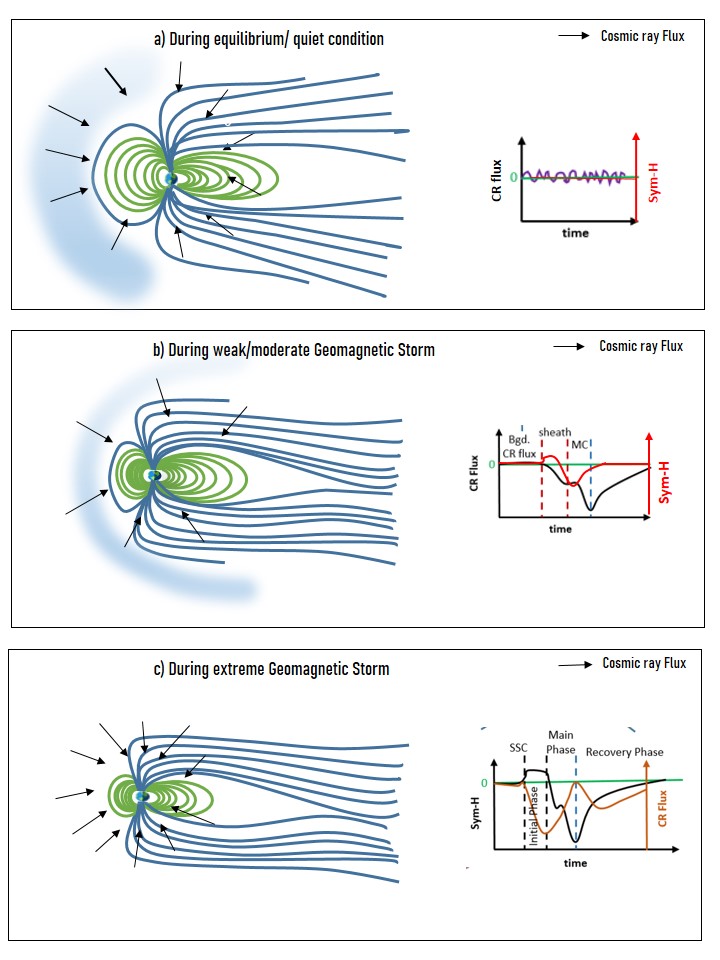}
	\caption{Schematic demonstration of Earth magnetosphere and cosmic-ray bombardment during quiet /equilibrium, weak /moderate storm, and extreme storm conditions. The typical cosmic ray flux and Sym-H index variations are displayed in the respective right panels.}
	\label{fig:cartoon}
\end{figure}

It is important to note that there is no cosmic-ray flux inside the CME while ejecting from the Sun. In general, the CME expands due to high internal magnetic pressure during propagation throughout the heliosphere. The magnetic barrier, convection, or scattering processes do not allow the comic ray to enter the ICME easily. The high-energy cosmic ray (depending on the rigidity) can drill the barrier, whereas the low and intermediate energy cosmic ray particles cannot penetrate the barrier. However, some cosmic ray flux slowly diffused (depending on the rigidity) inside the ICME. Therefore, when ICME with depleted cosmic ray flux passes over the Earth, ground-based neutron monitors generally show a decrease in their flux and are called FD phenomena. Besides, an additional shield layer exists for ground-based cosmic ray detectors, i.e., Earth magnetosphere. The magnetosphere also resists the cosmic ray flux depending on the rigidity of a particular location. The rigidity varies from low to maximum from pole to equator of the Earth. Thus, incoming cosmic ray particles can reach the particular location if and only if the energy of the incoming particle is greater than the rigidity of the location. Therefore, only a high-energy cosmic ray can enter the magnetosphere, but the lower energy cosmic ray flux is blocked. For example, a polar or high latitude station can measure cosmic rays with energy 0.1 GV; however, only the high energy ($> 14~GV$) cosmic ray flux can reach the NM detectors situated close to the equator.

Figure \ref{fig:cartoon} describes a schematic representation to explain the role of geomagnetic disturbance during FD phenomena. The top panel (Figure \ref{fig:cartoon}a) demonstrates a schematic diagram of the magnetosphere during the quiet condition. In this case, the Earth's magnetic field does not vary significantly. Moreover, as the cosmic rays are omnipresent, we observe a steady background cosmic ray flux variation.
Whereas Figure \ref{fig:cartoon}b and Figure \ref{fig:cartoon}c describe the situations when an ICME hit the Earth's magnetosphere, and causing moderate/extreme GS and FD phenomena. The middle panel indicated the typical magnetosphere's response during weak/moderate storms. The right side of this panel describes the large amplitude FD profile and small variation in the Sym-H index. The bottom panel shows the probable magnetosphere's response to the extreme geomagnetic storm. The right side of the panel shows the enhancement in cosmic ray flux during FD associated with the largest decrease of the geomagnetic field.

When an ICME interacts with the Earth's magnetosphere, sometimes it induces extreme geomagnetic storm, which results in the weakening of the Earth's magnetic field. This causes the reduction of magnetic rigidity significantly around the globe. The weakening of the Earth's shielding is observed in the Sym-H index data. Moreover, the decrease in the Sym-H index is an approximate measure of the intensity of the ring current generated field.  It implies that the magnetic rigidity of equatorial and low latitude neutron monitor stations should be decreased significantly during these situations.  Thus, more cosmic ray flux should enter the Earth's magnetosphere and reach the ground during the extreme GS. Their interaction with atmospheric molecules enhances the neutron flux. Therefore, we observed a significant gradual increase of high energy neutron monitors (situated in the low-latitude regions) flux compared to intermediate / low energy monitors. We believe that the decrease in magnetic rigidity for high altitude stations or near-polar stations will be low. Therefore, we cannot see a significant enhancement in high latitude compared to low latitude stations during the main phase of the geomagnetic storm. However, the gradual enhancement is observable for low rigidity stations as well.


In summary, we conclude that an ICME causes GS and FD phenomena. The step decrease in the FD profile is associated with the ICME sub-structures. However, ICME-magnetosphere interactions significantly weaken the magnetosphere shield of the Earth during extreme storm conditions, which decreases the rigidity cut-off globally. Therefore, we observe a local hump-like profile along with the step decrease in FD profile. It is important to note that lower latitude stations show higher enhancements. This implies that ring current dynamics (associated with the storm) play a significant role in cosmic ray modulations. We also proposed that it is essential to include the effect of magnetospheric field dynamics along with the external causing agent (like convection, diffusion, turbulence, etc.) while studying the FD profile. Therefore, to reproduce a complete FD profile, a new model needs to be developed, including this effect. A more detailed study is needed to explore how the magnetosphere plays a role in the FD profile.  It is also essential to explore why mid to high latitude NM stations do not show significant hump-like variation during extreme GS. 
The detailed investigation might put some light on the ICME magnetosphere coupling/interaction process.	



\section*{Acknowledgements}

We acknowledge the NMDB database (\url{www.nmdb.eu}) founded under the European Union's FP7 programme (contract no. 213007), and the PIs of all individual neutron monitors. We also acknowledge J.H. King, N. Papatashvilli, AdnetSystems, NASA GSFC and \url{https://cdaweb.gsfc.nasa.gov/}. Anil Raghav and his JRF Mr. Omkar Dhamane is supported by the Science \& Engineering Research Board (SERB) project file reference no. (CRG/2020/002314). We thanks Department of Physics, University of Mumbai for providing facilities to perform this work. ZS is supported by the Department of Science \& Technology (DST) government of India. We also thanks to the anonymous reviewer for giving valuable comments and suggestion which helped us to
improve the readability.

\section*{Data Availability}

The data used in this article are publicly available in the following
repository; (1) The world-wide neutron monitor archive data available at \url{https://www.nmdb.eu/nest/}
 and (2) Coordinated Data Analysis Web (CDAWeb) available
at \url{https://cdaweb.gsfc.nasa.gov/cgi-bin/eval2.cgi}.

\bibliographystyle{hapalike} 
\bibliography{FD}


\end{document}